\documentclass{article}
\usepackage{natbib}
\usepackage{graphicx,color}

\begin{document}
\newcommand\affiliation[1]{\gdef\@affiliation{#1}}
\gdef\@affiliation{}
\newcommand\etal{\mbox{\textit{et al.}}}
\relax

\title
{Investigation of the recombination of the retarded shell of
``born-again'' CSPNe by time-dependent radiative transfer models
\footnote{to appear in IAUS 283, proceedings of the IAU Symposium \protect{\newline}
\indent\ \ ``Planetary Nebulae: An Eye to the Future'', \protect{\newline}
\indent\ \ Eds.: A. Manchado, L. Stanghellini and D. Sch{\"o}nberner}
}

\author
{Antti Koskela$^1$, Silvia Dalnodar$^2$, Ralf Kissmann$^2$, Anita \\Reimer$^3$,
Alexander Ostermann$^1$ and Stefan Kimeswenger$^{2}$\footnote{presenting author, Stefan.Kimeswenger@uibk.ac.at}
}

\date{\scriptsize{$^1$Department of Mathematics, University Innsbruck, Technikerstrasse 13/7, \\A-6020 Innsbruck, Austria\\\smallskip
$^2$Institute of Astro- and Particle Physics, University Innsbruck, Technikerstrasse 25/8, \\A-6020 Innsbruck, Austria\\\smallskip
$^3$Institute of Theoretical Physics, University Innsbruck, Technikerstrasse 25/2, \\A-6020 Innsbruck, Austria\\}}

\maketitle

\begin{abstract}
A standard planetary nebula stays more than 10 000 years in the state of
a photoionized nebula. As long as the timescales of the most important
ionizing processes are much smaller, the ionization state can be
characterized by a static photoionization model and simulated with
codes like {\tt CLOUDY} (Ferland et al. 1998). When
the star exhibits a late Helium flash, however, its ionizing flux
stops within a very short period. The star then re-appears from its
opaque shell after a few years (or centuries) as a cold giant star
without any hard ionizing photons. Describing the physics of such
behavior requires a fully time-dependent radiative transfer
model. Pollacco (1999), Kerber
  \etal\ (1999) and Lechner \& Kimeswenger
  (2004) used data of the old nebulae around V605 Aql and
V4334 Sgr to derive a model of the pre-outburst state of the CSPN in a
static model. Their argument was the long recombination time scale for
such thin media. With regard to these models Sch{\"o}nberner
  (2008) critically raised the question whether a
significant change in the ionization state (and thus the spectrum) has
to be expected after a time of up to 80 years, and whether static
models are applicable at all.

\noindent{{\it Keywords:} {\tt radiation mechanisms: general, ISM: general, (ISM:) planetary nebulae: individual (V604 Aql, V4334 Sgr)}}

\end{abstract}

\begin{figure}[ht]
\begin{center}
\scalebox{0.49}{\includegraphics[angle=0]{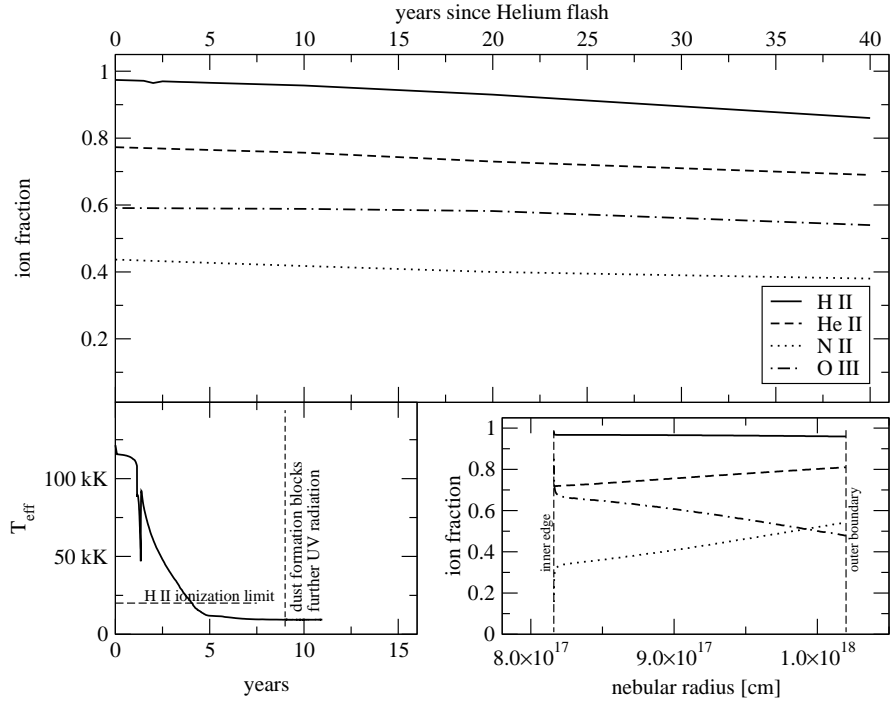}}
 \caption{The upper panel shows the resulting time evolution of the
 ionization state of the nebula. We show  only those ions which were
 used for photoionization analysis for the pre-flash white dwarfs.
 The state at the mean radius of the nebula is shown here. The solution
 of a model including the cooling luminous star while  returning to the
 AGB (lower left panel), and that completely without any ionizing source,
 differ by less than the line thickness. The lower right panel shows the
 spatial distribution after 20 years. Due to the low  filling factor, the
 time evolution turns out to be similar for all radii. In contrast to the
 simulation by  Binette \etal\ (2003), no outer recombination rim is formed.}

\end{center}
\end{figure}

We use the
  transport equation formalism presented in the appendix of
Binette \etal\ (2003) to describe the evolution of the
  nebula after the rapid change of the central star. As extensions to their model we also take {\tt He} and   the
     {\tt CNO} elements into account, both as source of free electrons
     and as absorbing species. Also, we use a
     realistic stellar photosphere model for the input
     radiation field, and, as we investigate the evolution for a large
     change of the input radiation field, we do not use their
     monochromatic approach of a single mean frequency.  The system
     is solved using a classical splitting scheme
     and the pressure and temperature are calculated using an explicit
     solver.  The physical parameters were taken from
     Osterbrock \& Ferland (2006) and Gnat
       \& Sternberg (2007). The initial conditions were
     calculated by {\tt CLOUDY}. For the evolution of the
     ``born-again'' star in the Hertzsprung-Russell-Diagramm the
     tracks as used in van Hoof \etal\ (2007) were
     applied.

The nebulae of the "born-again" central stars recombine
extremely slowly (see Fig. 1). The Hydrogen and Helium timescales
dominate the whole process. Pollacco (1999) raised
the question about the Oxygen recombination time scale. As single
species it would be only about one decade. The coupling of the Oxygen
processes with that of the Helium, however, slows down the
recombination, as long as {\tt He II} exists, to a rate even below
that of Hydrogen.

\noindent The ionization of the luminous, but much colder
star during the transit back to the AGB phase, stated by
Sch{\"o}nberner (2008) as possible major
mechanism for reionization, is negligibly small.  Thus the
previous results of the pre-flash state of V4224 Sgr (Pollacco
  1999 and Kerber \etal\ 1999) and V605
Aql (Lechner \& Kimeswenger 2004) are reliable.

\smallskip

\noindent{\sl Acknowledgements:\\}{\small A.K. and S.D. are funded by Austrian Science Fund (FWF) DK+ project {\sl Computational Interdisciplinary Modeling}, W1227}
\vspace{5mm}

%%\begin{thebibliography}{}
%%\bibitem[Binette \etal\ (2003)]{binette03}
\centerline{\bf References:}
\noindent{Binette, L., Ferruit, P., Steffen, W., \& Raga,
A.C.} 2003, \textit{Rev. Mexicana A\&A}, 39, 55
\smallskip

%\bibitem[Ferland \etal\ (1998)]{ferland98}
\noindent{Ferland, G.J., Korista, K.T., Verner,
D.A., Ferguson, J.W., Kingdon, J.B., \& Verner, E.M.}
1998, \textit{PASP}, 110, 761
\smallskip

%\bibitem[Gnat \& Sternberg (2007)]{gnat07}
\noindent{Gnat, O., \& Sternberg, A.} 2007, \textit{ApJS}, 168, 213
\smallskip

%\bibitem[Kerber \etal\ (1999)]{kerber99}
\noindent{Kerber, F., K{\"o}ppen, J., Roth, M., \& Trager, S.C.} 1999, \textit{A\&A}, 344, L79
\smallskip

%\bibitem[Lechner \& Kimeswenger (2004)]{lechner04}
\noindent{Lechner, M.F.M., \& Kimeswenger, S.} 2004, \textit{A\&A}, 426, 145
\smallskip

%\bibitem[Osterbrock \& Ferland (2006)]{osterbrock06}
\noindent{Osterbrock, D.E., \& Ferland, G.J.} 2006,
\textit{Astrophysics of gaseous nebulae and active galactic nuclei}, Sausalito, CA: University Science Books
\smallskip

%\bibitem[Pollacco (1999)]{pollacco99}
\noindent{Pollacco, D.} 1999, \textit{MNRAS}, 304, 127,
\smallskip

%\bibitem[Sch{\"o}nberner (2008)]{schoenberner08}
\noindent{Sch{\"o}nberner, D.} 2008,
in: K. Werner \& T. Rauch (eds.)
\textit{Hydrogen-Deficient Stars}, Proc. ASP Conference Series, Vol. 391 (San Francisco: ASP), p.\ 139
\smallskip

%\bibitem[van Hoof \etal\ (2007)]{vanhoof07}
\noindent{van Hoof, P.A.M., Hajduk, M., Zijlstra, A.A., Herwig, F., Evans, A., van de Steene, G.C., Kimeswenger, S., Kerber, F., \& Eyres, S.P.S.}
2007, \textit{A\&A} 471, L9
\smallskip

%\end{thebibliography}

\end{document}